\documentclass[preprints,article,accept,moreauthors,pdftex]{Definitions/mdpi} 

\firstpage{1} 
\pubvolume{1}
\issuenum{1}
\articlenumber{0}
\pubyear{2021}
\copyrightyear{2021}
\externaleditor{Academic Editor: Dimitris M. Christodoulou} 
\datereceived{17 August 2021} 
\dateaccepted{13 October 2021} 
\datepublished{} 
\hreflink{https://doi.org/} 
\pdfoutput=1

\newcommand{\ergs}{{\rm \,erg\,s^{-1}}}
\newcommand{\msun}{M_{\odot}}
\newcommand{\mbh}{{M_{\rm BH}}}
\newcommand{\ledd}{{L_{\rm Edd}}}
\newcommand{\ldisk}{{L_{\rm disk}}}
\newcommand{\lx}{{L_{\rm X}}}
\newcommand{\lr}{{L_{\rm R}}}
\newcommand{\pjet}{{P_{\rm jet}}}

\Title{The Origin of Radio Emission in Black Hole X-ray Binaries}

\TitleCitation{The Origin of Radio Emission in Black Hole X-ray Binaries}

\Author{Xiang Liu $^{1,2,3,}$*\orcidA{}, Ning Chang $^{1,4}$\orcidB{}, Xin Wang $^{1,4}$\orcidC{} and Qi Yuan $^{1,4}$}

\AuthorNames{Xiang Liu, Ning Chang, Xin Wang and Qi Yuan}

\AuthorCitation{Liu, X.; Chang, N.;\linebreak Wang, X.; Yuan, Q.}

\address{%
$^{1}$ \quad Xinjiang Astronomical Observatory, Chinese Academy of Sciences, 150 Science 1-Street,\linebreak Urumqi 830011, China; changning@xao.ac.cn (N.C.); wangxin2019@xao.ac.cn (X.W.); yuanqi@xao.ac.cn (Q.Y.)\\
$^{2}$ \quad Key Laboratory of Radio Astronomy, Chinese Academy of Sciences, Nanjing  210008, China\\
$^{3}$ \quad Key Laboratory of Radio Astrophysics in Xinjiang Uygur Autonomous Region, Urumqi 830011, China\\
$^{4}$ \quad School of Astronomy and Space Science, University of Chinese Academy of Sciences, Beijing 100049, China\\}

\corres{Correspondence: liux@xao.ac.cn}

\abstract{We studied the relation of accretion-jet power and disk luminosity, especially the jet efficiencies and disk radiative efficiencies for different accretion disks as well as black hole (BH) spin, in order to explore the origin of radio emission in black hole X-ray binaries (BHXBs). We found that jet efficiency increases more rapidly (efficient) than the nearly constant disk radiative efficiency for thin disk component in high accretion regime, which could account for the steep track ($\mu>1$) in the observed radio and X-ray luminosity relations ($\lr\propto L_{\rm X}^{\mu}$), but the thin disk component may not be able to explain the standard track ($\mu\approx 0.6$) in the BHXBs. For hot accretion flows (HAF), the resulting jet efficiency changes along with the large range of accretions from quiescent state to nearly Eddington state, which could account for the standard track in the BHXBs. The BH spin-jet is discussed for the magnetic arrested disk (MAD) state; in this state, the spin-jet power might contribute to a linear correlation between jet power and mass accretion rate for a given source. More accurate observations are required to test the results.}


\begin{document}

\section{Introduction}

Among compact binary systems, those that consist of a main sequence or evolved star (S) and compact objects such as a black hole (BH), a neutron star (NS) or a white dwarf (WD) are of special interests. Electromagnetic emission is often observed in  close BH-S, NS-S, and WD-S binaries, when the stellar matter is accreting onto the compact object. For the purpose of understanding the origin of the radio emission, in this paper we investigate the correlation between radio jet and other properties in black hole X-ray binaries (BHXBs). We chose BH systems because there are only two parameters (mass and spin) for a  BH, and there is another advantage, i.e., BH has no solid surface (the situation is more complicated for close NS-S, or WD-S binaries due to the strong magnetic field structure and the hard surface of compact object). Over the past decades, we have archived a sufficient amount of data on the radio emission from BHXBs. In this work, we will briefly introduce the radio properties of those BHXBs and investigate, from a theoretical point of view, the plausible/sound models for their radio origin.
 
\section{The Radio Emission of BHXB}\label{sec2}

From X-ray observations, there are about 60 transients classified as BH candidates. Among them, about 20 are dynamically confirmed as BHXBs ({\url{https://www.astro.puc.cl/BlackCAT/transients.php}}, accessed on 5 August 2021) \cite{corra16}. The X-ray properties of BHXBs can be studied through the intensity-hardness diagram and the so-called `q-diagram' (e.g., \cite{Fender14,Carotenuto2021}), where a flare/outburst evolves from a low luminosity-hard spectral state to a high luminosity-soft spectral state. A radio jet often appears in the low-hard state and sometimes evolves to intermediate soft state \cite{Fender14,Carotenuto2021}. The correlation between radio and X-ray luminosities is intensively investigated over the past decades. With the expression of observed radio and X-ray luminosity relations ($\lr\propto \lx^\mu$), two correlations have been reported. One is normal/standard, with power-law index of $\mu$$\sim$$0.6$ (also called `standard track', \cite{Fender04}), and the other is a steep one observed in some sources, with $\mu$$\sim$$1.4$   (also called `outliers track' or `steep track', e.g., \cite{Gallo12,corbel13}). We note that most sources cannot be analyzed in a timing manner, i.e., few sources have sufficient long-term monitoring in both radio and X-ray (few exceptions are, e.g., GX 339-4). One practical method will then be to study the correlation models with the data collected from different BHXBs.

Below, we briefly summarize the radio properties of BHXBs according to their radio phases from flare/outburst state to quiescent state. In the flare state, for example, a highly relativistic and confined ($\sim$$0.5^{\circ}$) jet is observed in MAXI J1820+070, which carries a significant amount of power away from the system (equivalent to $0.6\times L_{1-100~\text{keV}}$) \cite{Tetarenko21}. The jet shows a large bulk Lorenz factor of 6.8 with a high Eddington ratio of $\sim$$0.1$, and its total radio/sub-mm flux density peaks at roughly $>$$1000$ GHz \cite{Tetarenko21,Connors19}. The emission at different wavebands correlates with each other, with time-lags ranging from hundreds of milliseconds between the X-ray/optical bands to minutes between the radio/sub-mm bands \cite{Tetarenko21}. The derived magnetic field from model fitting is about $10^{4}$ Gauss in the jet base region \cite{Tetarenko21}. The well-known microquasars SS433 and GRS 1915+105 also demonstrated relativistic jets with the large jet-viewing angles with respect to our line of sight. The long-term radio and X-ray observations revealed quasi-periodic (about 200--300 days) flux variations in both radio and X-ray light curves of GRS 1915+105 \cite{Motta21}, which may imply quasi-periodic outbursts and accretion state changes. In SS433, such accretion state change is not observed, possibly because  disk emission is hidden from us in X-rays \cite{Paragi12}.

In the quiescent X-ray state, radio emission has been also detected in some BHXBs. Gallo et al. \cite{Gallo19} report on the Atacama Large Millimeter Array (ALMA) continuum observations of the black hole X-ray binary A0620-00 at an X-ray luminosity nine orders of magnitude sub-Eddington, the system was significantly detected at 98 GHz ($44\pm7~\upmu$Jy) \cite{Tremou20}. The recently obtained detections of the black hole X-ray binary GX 339-4 in quiescence state using the Meer Karoo Array Telescope (MeerKAT) radio telescope and Swift X-ray Telescope instrument on board the Neil Gehrels Swift Observatory reached the lower end of the radio-X-ray correlation.

There is also a `radio-faint' phase as noted in \cite{Plotkin17}. Swift J1753.5-0127 in the hard state emitted less radio emission from its jet than what expected from the standard track of the radio-X-ray correlation. Instead, it seems to follow the `radio-faint' phase, which is best observed in another BHXB H1743-322. In H1743-322, there is an unexpectedly horizontal branch in the $\lr\text{--}\lx$ plane when the X-ray luminosity is located between $\lx$$\sim$$4\times 10^{36}\ergs$ and $\lx$$\sim$$10^{35}\ergs$, and when it is fainter than $\lx$$\sim$$10^{35}\ergs$ it rejoined the `standard track' \cite{Jonker10,Criat11,Plotkin17}. The J1753.5-0127 has a relatively flat spectrum of $F_{\nu}\propto \nu^{0.3}$ , but it is still not detected with the Very Long Baseline Array (VLBA) above a 5$\sigma$ upper limit of 0.16~mJy/beam \cite{Tomsick15}.

\section{The Origin of Radio Emission}

As introduced above, the BHXBs can have relativistic jets (e.g., in microquasars), transient jets (in flare or mini-flares), and weak radio emission (in `radio-faint' and quiescent state). There are mainly two observational tracks/paths in the $\lr\text{--}\lx$ plane, the `standard track' ($\mu$$\sim$$0.6$) and the `steep track' ($\mu$$\sim$$1.4$). Radio emission can be powered by the accretion process and the extraction of BH spin energy. Accretion power can be scaled by the Eddington luminosity, $\ledd=1.3\times 10^{38}(\mbh/\msun)\ergs$. Weakly accreting systems, such as BHXBs with X-ray luminosities $\lx\lesssim10^{37}\ergs$ (or equivalently, Eddington ratio $\lx/\ledd\lesssim0.01$), cover both the hard X-ray spectral state ($10^{-5}\lesssim \lx/\ledd \lesssim 10^{-2}$, \cite{Remillard06}) and the quiescence state ($\lx/\ledd\lesssim10^{-5}$, \cite{Plotkin13}). As noted in \cite{Plotkin17}, when weakly accreting BHXBs brighten on day-to-week timescales, they can trace distinct paths in the $\lr\text{--}\lx$ plane (e.g., \cite{corbel13,Gallo14}). From a theoretical point of view, the radio emission is partially self-absorbed synchrotron radiation from a steady, unresolved, and flat-spectrum jet \cite{bland79,Fender01}, while the X-rays probe the inner regions of the accretion flow \cite{Fender14}. In this case, the presence of correlated radio and X-ray variability suggests a physical connection between the jet and the emission regions closest to the black hole \cite{Heinz03}.

\subsection{Accretion Disks/States and Jets}

In the following, we investigate the disk-jet coupling and the variations of the jet efficiency and disk radiative efficiency with increasing disk luminosity (or accretion rate). Then, we introduce different accretion disks and comment on possible connections between the efficiency of radio emission and accretion states of BHXBs.

The disk-jet coupling can be a consequence of the connection between the jet power $P_{\rm jet}^{\rm acc}=\eta \dot{M}c^{2}$ and accretion disk luminosity $\ldisk=\varepsilon \dot{M}c^{2}$. Thus, we have the following.
\begin{equation}
\label{eq1}
\pjet = (\eta/\varepsilon)\ldisk.
\end{equation}

In reality, most of the jet power $\pjet$ is in the form of mechanical energy (e.g., the work on the environment) but not radio luminosity $\lr$. From a sample of radio galaxies, Willott et al. \cite{willott99} found an empirical relationship between the time-averaged jet power and the monochromatic radio luminosity at 151 MHz, i.e., $\pjet=3.4\times 10^{20}\times f^{3/2}\times (\frac{L_{\rm 151}}{W Hz^{-1}})^{6/7}(\ergs)$, where parameter $f$ accounts for the theoretical uncertainties. We adopt this formula to calculate the jet power with $f = 10$ \cite{Godfrey13}, which can be read as follows.
\begin{equation}
\label{eq2}
\pjet\approx c_{1}L_{\rm R}^{6/7} \ergs,
\end{equation}
here, $c_{1} = 5.8\times10^8$. In this expression, we take a $\nu^{-0.8}$ spectrum in radio (between 151 MHz and 5 GHz) and define $L_{R}$ as the radio luminosity at 5 GHz. Note that this relation was derived from a sample of radio galaxies rather than BHXBs; thus, we caution that the application to BHXBs may be risky, and the jets in BHXBs are more compact, but our results do not change significantly if we used a linear or slightly lower index than (6/7) in the relation.

From Equations (\ref{eq1}) and (\ref{eq2}), we have the following.
\begin{equation}
\label{eq3}
\eta/\varepsilon\propto L_{\rm R}^{6/7}/\ldisk.
\end{equation}

For the BHXBs, disk luminosity can be approximately scaled to the X-ray luminosity in the inner accretion region where the radio jet is produced, as $\ldisk\propto\lx$, because X-ray luminosity is dominant in the spectral energy distribution of the inner disk \cite{Yuan14} (see next parts for more discussions). With this approximation, the `standard track' of BHXBs ($\lr\propto L_{\rm X}^{\mu}$ and $\mu$$\sim$$0.6$) will imply $\eta/\varepsilon\propto L_{\rm X}^{0.6\times(6/7)}/\lx\propto L_{\rm X}^{-0.5}$, while the steep track of BHXBs with $\mu$$\sim$$1.4$ will imply $\eta/\varepsilon\propto L_{\rm X}^{1.4\times(6/7)}/\lx\propto L_{\rm X}^{0.2}$. This suggests that the $\eta/\varepsilon$ can be assumed as a power-law function of the disk X-ray luminosity, with the index of q. In other words, we have the following:

\begin{equation}
\label{eq4}
\eta/\varepsilon\approx c_{2} L^{q}_{\rm disk},
\end{equation}
where the $q$ reflects the dependence of the combined factor $\eta/\varepsilon$ on the disk luminosity or accretion rate.

From the Equations (\ref{eq1}) and (\ref{eq4}), we have the following:
\begin{equation}
\label{eq5}
\pjet=(\eta/\varepsilon) \ldisk=c_{2}L_{\rm disk}^{\mu'}=c_{2}L_{\rm disk}^{1+q}=c_{3}(\lambda \mbh)^{1+q},
\end{equation}
where $c_{2}$ and $c_{3}$ are the two constants, and $\mu'=1+q$, $\lambda=\ldisk/\ledd$ is the Eddington ratio.

Equation (\ref{eq4}) can be applied to study the non-linear correlation between jet power and disk luminosity in BHXBs (negative $q$ when $\mu'<1$ and positive $q$ when $\mu'>1$). The dependence of these efficiencies on mass accretion rate, key for our understanding of radio and X-ray activities and radio-X-ray correlation, is poorly investigated. In the following, we will try to investigate these efficiencies for different accretion disks in the BHXBs and comment on the origin of radio emission in BHXBs.

\subsubsection{Geometric Thin/Cold Disk}

The SSD (Shakura--Sunyaev Disk, \cite{Shakura73}) is a geometrically thin but optically thick disk. It emits thermal blackbody-like radiation. Many accreting black holes have been successfully modeled as thin disks, e.g., luminous active galactic nuclei (AGN) and the BHXBs in the thermal state (e.g., \cite{Remillard06}). For example, the persistent BHXB source GRS 1915+105 with relativistic super-luminal radio ejections is believed to accrete erratically close to the Eddington limit \cite{Mirabel1994}. Over the past three decades, GRS 1915+105 follows the `steep track' in the $\lr\text{--}\lx$ plane \cite{Motta21}.

In the SSD, the disk radiative efficiency $\varepsilon$ varies in the range of 0.057--0.43, depending on the BH spin \cite{NovikovT,Narayan21}. The $\varepsilon$ of SSD has a weak or no dependence on disk luminosity, so $\varepsilon_{\rm SSD}=0.1$ is widely adopted \cite{Shakura73,NovikovT}. In the upper panel of Figure \ref{fig1}, we have fitted the radio and X-ray (1--10 keV) data from \cite{Espinasse18}. We find $\mu = 0.64\pm0.02$ and $1.38\pm0.14$ for the standard track (filled circles) and the steep track (open triangles), respectively. If we assume that the accretion is cold-disk-like and follows $\varepsilon_{\rm SSD}=0.1$ efficiency profile, then we can calculate the jet production efficiency $\eta$ based on Equations (\ref{eq1}) and (\ref{eq2}) where the disk luminosity is approximated as $\ldisk\approx3\lx$ in a hard state (Fu-Guo Xie, private communication, from modeling the SED of BHXB, e.g., \cite{Yuan05,Yuan14}).

The jet production efficiency for different types of tracks is shown in the lower panel. From this plot, we find that for those of 'standard track' data (filled circles in the upper panel), the jet production efficiency $\eta$ for the SSD case decreases as accretion rate (or X-ray luminosity) increases. This is shown by the orange dotted line. On the other hand, for those of `steep track' data (open triangles in the upper panel), we find that the jet production efficiency has a positive correlation with the accretion rate (or X-ray luminosity). This is shown by the blue dotted line. Due to the fact that, in BHXBs, the SSD mainly works at the thermal state and high accretion rates (bright $\lx$ regime; e.g., see Figure 7 in \cite{narayan1998}), the SSD may not be responsible for the standard track, which is extending to the low $\lx$. On the other hand, the $\eta$ for the SSD case is derived from the steep track data, with $\lx/\ledd>0.001$, where the SSD works. It implies that the SSD (or a thin-disk component in hot accretion flows, see discussion below) may be responsible for radio emission in the steep track.
\subsubsection{Geometric Thick/Hot Acrretion Flow (HAF)}

Another class is hot accretion flow (HAF), which is conventionally named as advection-dominated accretion flow (ADAF, \cite{Narayan94,Yuan14}). Due to energy/entropy advection, HAF is hot (thus, geometrically thick), and it is usually radiativly inefficient. The radiative efficiency of a HAF increases with increasing mass accretion rate and can even be comparable to that of an SSD \cite{Xie12}. {This is especially the case for the bright two-phase medium with cold dense clumps embedded in hot gas in the so-called luminous hot accretion flow (LHAF) regime~\cite{Xie12,Yuan14}.}

\begin{figure}[H]
	\includegraphics[width=9cm]{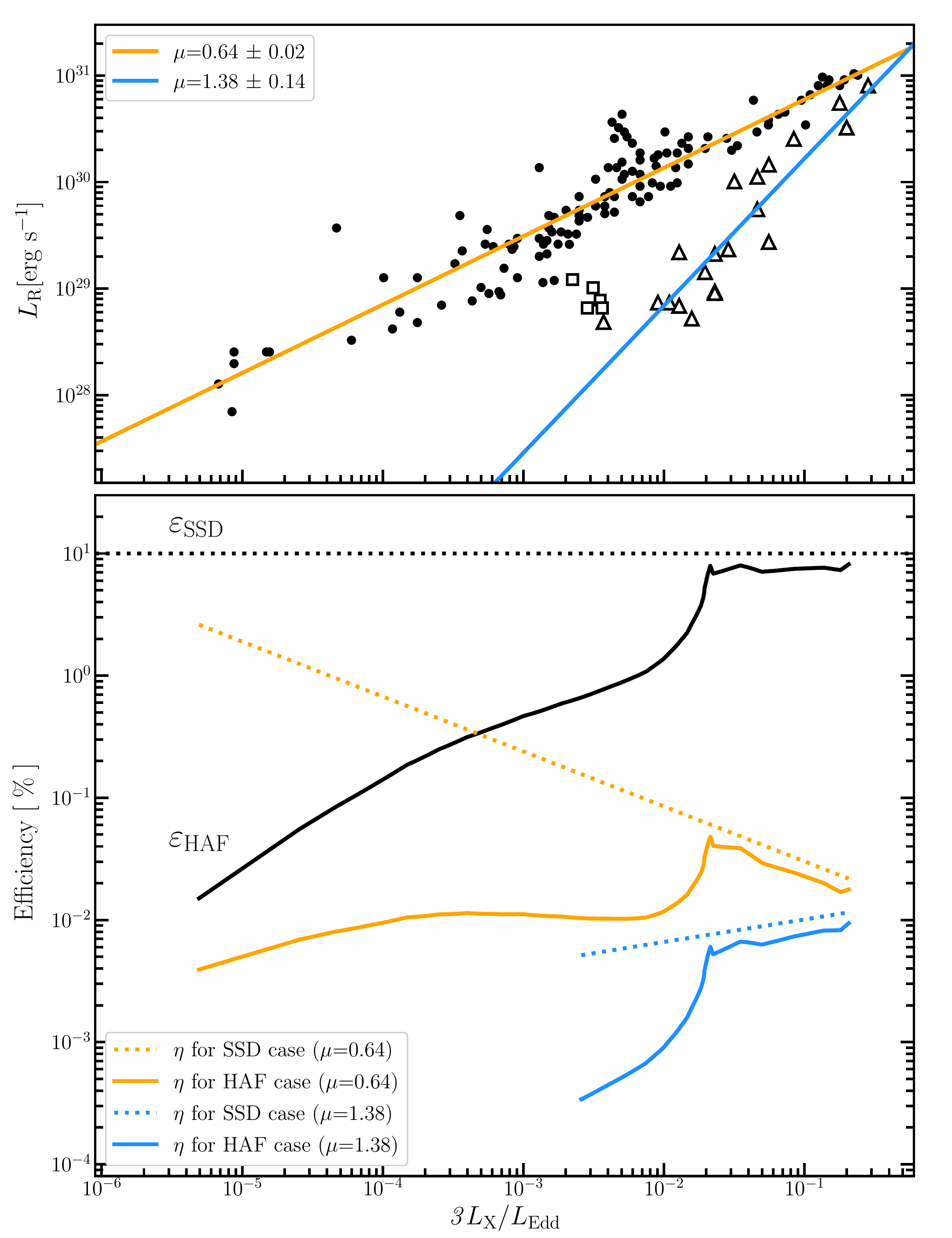}
	\caption{The upper panel shows the radio-X-ray correlation of BHXBs with data taken from \cite{Espinasse18}. In this plot, the filled circles are those of the standard track, while the open triangles are those of the steep track. The fitting results are also shown here, where we assume BH mass is $10\msun$. The lower panel shows the jet production efficiency $\eta$ based on the radiative efficiency $\varepsilon$ of various accretion flow models (dotted lines for $\eta$ of SSD and solid lines for $\eta$ of the hot accretion flow, with different colors corresponding to the different slopes $\mu$). For the sake of completeness, we also show in the lower panel the radiative efficiency of SSD and HAF taken from \cite{Xie12} by the black-dotted and black-solid lines,~respectively.}
	\label{fig1}
\end{figure}

\begin{figure}[H]
	\includegraphics[width=12cm]{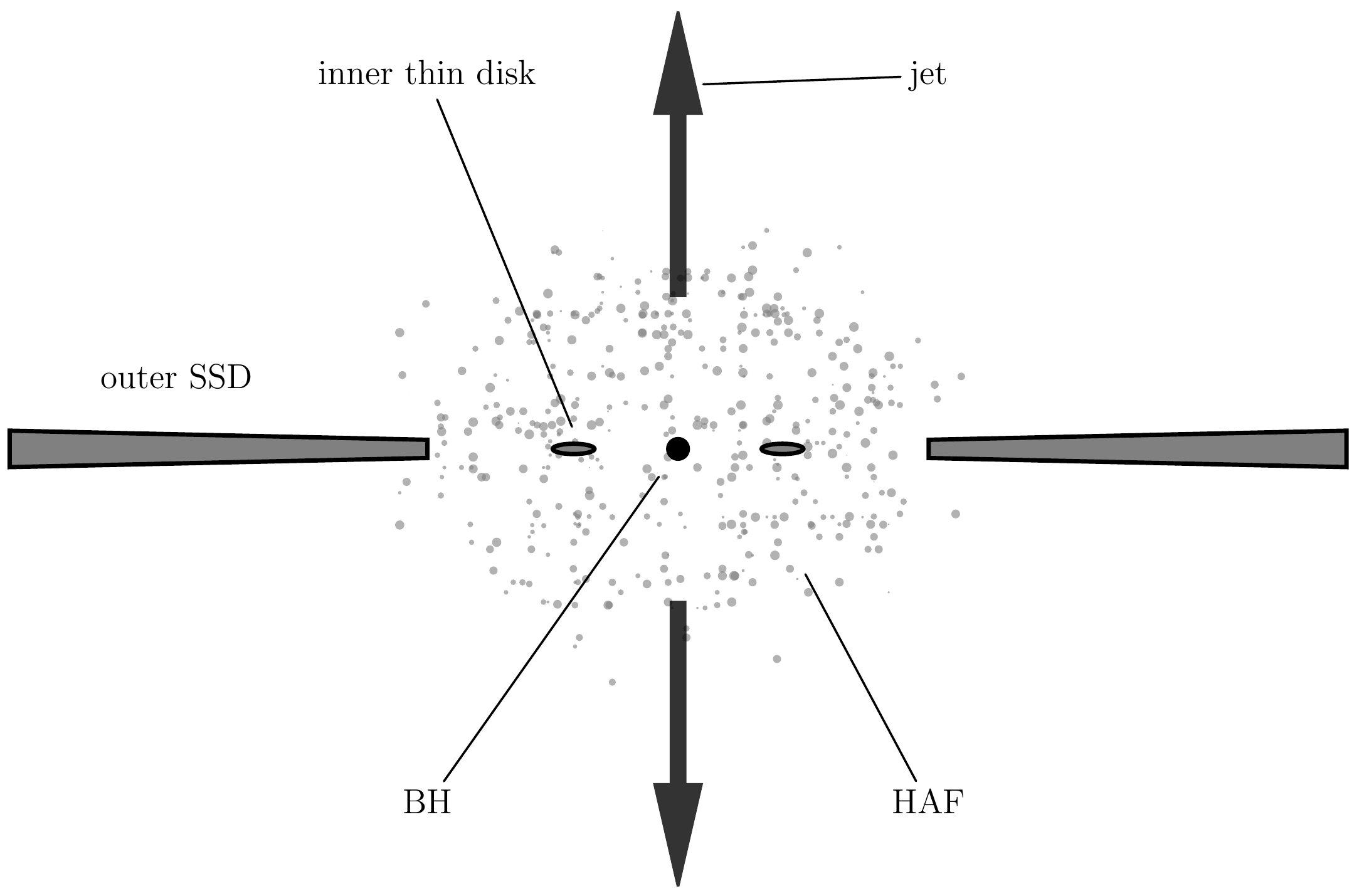}
	\caption{The schematic plot for a truncated disk with outer SSD and inside hot accretion flow (HAF) and the inner thin-disk component, as well as jets and central black hole.}
	\label{fig2}
\end{figure}

In fact, numerical calculations by \cite{Xie12} found that $\varepsilon$ increases rapidly with increasing accretion rate in the LHAF regime. It is suggested that the BHXBs may be more likely in the luminous hot accretion state than in normal hot accretion state in outbursts \cite{Yuan14}. With the $\varepsilon$ profile from \cite{Xie12}, a specific case with the electron viscous heating parameter $\delta=10^{-2}$ shown by the black solid line in the lower panel of Figure \ref{fig1}, we can also calculate the jet production efficiency. The results are shown in the lower panel of Figure \ref{fig1}. We found that because the radiative efficiency has a strong dependence on mass accretion rate now, the outcome jet production efficiency will also have a strong dependence on mass accretion rate. The standard track shows a stable $\eta$ at low $\lx$ regime. Interestingly, as noted in \cite{XieY16}, if we increase the viscous parameter (which helps to transport the angular momentum of accreting gas), then the ADAF/LHAF branch can extend to much higher $\lx$. In this case, the jet from the ADAF/LHAF branch may explain the standard track~\cite{XieY16}. The steep track, with $\lx/\ledd>0.001$, on the other hand may have two-phase medium with cold dense clumps embedded in hot gas \cite{Yuan14}, as its central engine. In this accretion mode, the radiative efficiency will be nearly constant and almost independent of X-ray luminosity (and accretion rate), which is similar to that of a thin-disk component as discussed above. The thin disk component may be dropped from outer SSD or be locally collapsed/condensed from the HAF \cite{Meyer-Hofmeister2014}, ({also see the schematic plot Figure \ref{fig2}).}

We note that the hard (and quiescent) state of BHXBs is successfully understood under the so-called truncated disk-jet model. In this model, the outer SSD is truncated at a certain radius, inside of which it is replaced by HAF. The jet can launch from a region very close to BH. In this model, it explains the key observational fact that there exist a thermal component (by outer SSD) and a power-law tail (by HAF) in its X-ray emission. Moreover, the truncation radius is also found to move inward as the system brightens (e.g., \cite{Yuan14}).

We caution that the origin of radio emission may have alternative explanations because the radio emission in the hard (and quiescent) state mostly cannot be resolved in BHXBs. In addition to jet,  the magnetic dissipation in the corona (or outflow) may also introduce a noticeable radio emission \cite{Laor08}. Indeed, the magnetic field strength near BH is fairly strong, i.e., modeling of observational data of both A0620-00 and XTE J1118+480 reports a value around 2000 Gauss at a radius of 10 $r_{g}$ ($r_{g}$ is the gravitational radius of a BH, see e.g.,~\cite{Wallace21}).

\subsubsection{Unstable Inner Disk/Discrete Accretion}

With the exception of continuous (relatively longer timescale) inner accretion flow, the mini-outbursts in BHXBs can also be understood by discrete accretions and/or unstable inner accretion disk. This model is applied to the three mini-outbursts of XTE J1550-564 by~\cite{dong21}. Xie et al. \cite{Xie20} studied the 2015--2016 mini-outbursts of GRS 1739-278. They found that the radio spectral index of the mini-outbursts is 0.28 $\pm$ 0.17 with JVLA (Karl G. Jansky Very Large Array) and observed a flat radio-X-ray correlation with $\mu$$\sim$$0.16$. The flat correlation is also found in the regime of the transition between the standard track and the steep track from some other sources, e.g., the five open squares in Figure \ref{fig1}  (see also \cite{Criat11,Carotenuto21}). Moreover, the mini-outbursts of Swift J1753.5-0127 follow two distinctive tracks \cite{Plotkin17}, i.e., a `standard track' but with lower normalization when $\lx\lesssim 10^{36}\ergs$, and a steep one when $\lx\gtrsim10^{36}\ergs$. Such behaviour is similar to the two tracks observed in main outbursts. The physical reason may be that the slope (the $q$ in Equation (\ref{eq4})) of $\eta/\varepsilon$ changes with accretion rates in mini-outbursts, where  the mini-outbursts may suffer lower viscous parameter $\alpha$ as suggested in \cite{Xie20}, as $\alpha$ controls the critical accretion rate (and luminosity) of different accretion models.

\subsection{BH Spin Contribution}

The BH spin may also play a role in determining jet power (radio emission,~\cite{Steiner13}), although more data are required to confirm this. For high-spin BHs, the Blandford--Znajek (BZ) mechanism \cite{bland77} could contribute a significant part of radio emission, which may have a different $\lr\text{--}\lx$ correlation from accretion-dominated jet from the Blandford--Payne (BP) mechanism \cite{bland82}. There is now a growing number of BHXBs that have spin measurements (see e.g., \cite{Reynolds21} for latest review) where the X-ray broad iron reflection lines and continuum fitting play the leading roles. For example, a near-extreme Kerr BH with a spin parameter $a_{*}\gtrsim 0.98$ is derived in Cygnus X-1 by \cite{Gou14} based on the continuum-fitting method. Considering the BZ mechanism, the jet power in Cygnus X-1 may be higher than normal. We will discuss the BZ-jet in the next section.

\section{Discussion}

The production of radio emission in both the BHXBs and AGN is coupled with the accretion state, the mass, and spin of BH. We observe a scale-invariant property (BH fundamental plane) of BH activity \cite{Merloni03,Falcke04,liu16} among BHXBs (stellar BH with $\mbh$$\sim$$3\text{--}20\msun$, typically $\mbh$$\sim$$10\msun$) and AGNs (supermassive BH with $\mbh >10^{6}\msun$). The timescale of an outburst in BHXBs ranges from days to years, and the $\mbh$ scaled up to the timescale of millions of years in AGNs, as suggested from AGN duty cycle (e.g., \cite{liu20,Baldi2021}). One exception is the so-called 'changing-look' AGN (e.g., \cite{Feng21}), and its state change timescale can be as low as months.

In this work, we consider the radiative efficiency from various accretion models. For the standard SSD, we adopt a $\dot{M}$-independent radiative efficiency $\varepsilon_{\rm SSD}=0.1$. For the hot accretion flow, we follow \cite{Xie12} for radiative efficiency. The hot accretion flow investigated by \cite{Xie12} unified three sub-types according to the mass accretion rate, i.e., the traditional ADAF at low $\dot{M}$, a LHAF at higher $\dot{M}$ (e.g., \cite{YuanZ04}), and a two-phase medium with cold dense clumps embedded in hot gas when $\dot{M}$ is fairly high \cite{Yuan14}. This hot accretion flow may be surrounded by a truncated cold SSD, and the transition may possibly be a result  of evaporation (from SSD) processes (e.g., \cite{Yuan14}). We note that in the inner hot accretion flows there may exist cold gas components (probably dropped from outer SSD), which gradually moved close to the BH in a spiral manner, similar to the tidal disruption event. The entire picture may explain both the main outburst and mini-outburst (from debris of the cold gas~components).

We note that the jet-viewing angle (between the approaching jet direction and our line of sight) effect has not been considered in this work. Most of the jet viewing angles in BHXBs are estimated to be >$20^\circ$ with large uncertainty (e.g., \cite{motta2018}), among which the relativistic beaming effect would be moderate. From a small sample of BHXBs, Motta et al. \cite{motta2018} found that the radio loudness of the BHXBs might be an inclination effect, which may suggest that the difference in $\lr\text{--}\lx$ tracks might be a geometric effect due to the different inclination angles. However, when accurate inclination angles for a large sample of BHXBs are obtained, one may be able to analyze whether a continuous change of viewing angle from $0\text{--}90^\circ$ can result in two distinct tracks with the slopes of 0.6 and 1.4. {Furthermore, an extremely edge-on accretion disk will obscure the X-rays from the inner disk, which will result in the underestimation of X-ray luminosity or even almost non-detection such as in SS433 mentioned in Section \ref{sec2}.}
	

As first proposed in the Blandford--Znajek (BZ, \cite{bland77}) jet model, with the help of a global ordered magnetic field, the BH spin energy can be extracted to launch a powerful relativistic jet. For comparison, other mechanisms rely on the extraction from the accretion disk, e.g., the case of the Blandford--Payne \cite{bland82} jet model, which is also named as disk-jet (see~\cite{Yuan14} for the clarification). The BZ jet is naturally produced and is the most powerful when the accretion flow supplied with a sufficient amount of magnetic flux and becomes a magnetic arrested accretion disk (MAD, \cite{Narayan03,Tchekhovskoy11}). The maximal/saturated magnetic flux threading the BH in the MAD state is $\Phi_{\rm MAD}\approx 1.5\times 10^{21} (\mbh/M_{\odot})^{3/2}(\dot{M}_{\rm BH}/\dot{M}_{\rm Edd})^{1/2}$ Gauss\,cm$^{2}$ \cite{Tchekhovskoy11,Tchekhovskoy12}, where $\dot{M}_{\rm Edd}$ is Eddington accretion rate. In the BZ jet, the jet power depends on the BH spin, the magnetic flux threading on BH, and the mass accretion rate~\cite{Tchekhovskoy11,Steiner13}. In other words, we have the following.

\begin{equation}
\label{eq6}
P_{\rm jet}^{\rm BZ}\approx 0.65a_{*}^{2}(1+0.85a_{*}^{2})(\Phi/\Phi_{\rm MAD})^{2}\dot{M}_{\rm BH}c^{2},
\end{equation}
here, $-1<a_{*}<1$ is the dimensionless BH spin. This expression works for the BZ jet for either the typical low-$\Phi$ hot accretion flow or the MAD one. The BZ jet disappears when $a_{*} = 0$ (no BH spin energy to extract). 

From the observational point of view, radio emission may originate from both disk-jet and BZ-jet \cite{Yuan14}, and the reported radio/X-ray correlation reflects the averaged properties, it is challenging to distinguish the contribution from these two types of jet. From $\sim$$20$ BHXBs with BH spin measurements \cite{Reynolds21}, we noticed that most sources ($\gtrsim$$2/3$) have fairly high BH spin (i.e., $a_{*} > 0.7$). Moreover, if some BHXBs in the hard state are MAD ($\Phi = \Phi_{\rm MAD}$; thus, $P_{\rm jet}^{\rm BZ}$ depends on BH spin and accretion rate only) with the argument by \cite{Narayan21} that MAD can easily be achieved, Equation (\ref{eq6}) may contribute to a linear correlation found between radio power and high accretion rate in MAXI J1348-630 \cite{Carotenuto21} given $P_{\rm jet}^{\rm BZ} \propto \dot{M}_{\rm BH} \propto \lx/\varepsilon$ (where the $\varepsilon$ approaches a constant in the high accretion regime in Figure \ref{fig1}) for the source. The BZ-jet (spin-jet) can be more magnetized and Poynting flux dominated than the accretion-jet, and the latter may be accompanied by outflow-winds. However, more accurate observations and models are required to investigate the spin-jet signatures for the lower magnetic flux regime $\Phi<\Phi_{\rm MAD}$.

\section{Summary and Outlook}

We studied the relation of accretion-jet power and disk luminosity, especially the jet efficiencies and disk radiative efficiencies for different accretion disks, as well as the BH spin-jet to explore the origin of radio emission in BHXBs. We find that jet efficiency increases more rapid (efficient) than the nearly constant disk radiative efficiency for thin disk component in the high accretion regime, which could account for the steep track in the $\lr\text{--}\lx$ plane, but the thin disk component may not be able to explain the standard track in the BHXBs. For the hot accretion flows (HAF), the resulting jet efficiency varies with the large range of accretions from quiescent to a nearly Eddington state, which could account for the standard track in BHXBs. The BH spin-jet is discussed for the MAD state in which the spin-jet power might contribute to a linear correlation between jet power and mass accretion rate for a given source. However, more accurate observations and models are required to investigate the spin-jet signatures in the sub-MAD regime.

For detailed analysis in future, the X-ray flux from the jet could be potentially separated from total X-ray flux in order to conduct a clean analysis of jet-disk coupling in BHXBs \cite{Debnath21}. The jet fraction in X-ray luminosity is quite large in \cite{Debnath21}; however, the modeling of the SED in BHXB XTE J1118+480 in the hard state shows a well-fitted X-ray emission with ADAF model, and the fitted jet contribution to X-ray luminosity is less than 10 percent \cite{Yuan05,Yuan14}. This difference should be investigated in future.

%

%

\vspace{6pt} 



\authorcontributions{Conceptualization, methodology, and writing---original draft, X.L.; formal analysis, visualization, and writing---review and editing, N.C., X.W., and Q.Y. All authors have read and agreed to the published version of the manuscript.}

\funding{This work is supported by the National Key R\&D Program of China under grant number 2018YFA0404602.}

\institutionalreview{Not applicable.}

\informedconsent{Not applicable.}

\dataavailability{Not applicable.}

\acknowledgments{We thank the referees for careful reports, Fu-Guo Xie for useful discussion on the manuscript, and Rob Fender for valuable communication on the data. We thank the support from the Key Laboratory of Radio Astronomy, Chinese Academy of Sciences, and the Key Laboratory of Radio Astrophysics in Xinjiang Uygur Autonomous Region of China.}

\conflictsofinterest{The authors declare no conflicts of interest.} 

\end{paracol}
\reftitle{References}


\begin{thebibliography}{999}
	
\bibitem[Corral-Santana et al.(2016)]{corra16}
Corral-Santana, J.M.; Casares, J.; Mu$\tilde{n}$oz-Darias, T.; Bauer, F.E.; Martinez-Pais, I.G.; Russell, D.M. BlackCAT: A catalogue of stellar-mass black holes in X-ray transients. {\em Astron. Astrophys.} {\bf 2016}, {\em 587}, 61.

\bibitem[Fender \& Gallo (2014)]{Fender14}
Fender, R.; Gallo E. An Overview of Jets and Outflows in Stellar Mass Black Holes. {\em Space Sci. Rev.} {\bf 2014}, {\em 183}, 23--337.

\bibitem[Carotenuto et al. (2021)]{Carotenuto2021}
Carotenuto, F.; Corbel, S.; Tremou, E.; Russell, T.D.; Tzioumis, A.; Fender, R.P.; Woudt, P.A.; Motta, S.E.; Miller-Jones, J.C.A.; Chauhan, J.; et al. The black hole transient MAXI J1348-630: Evolution of the compact and transient jets during its 2019/2020 outburst. {\em Mon. Not. R. Astron. Soc.} {\bf 2021}, {\em 504}, 444--468.

\bibitem[Fender et al. (2004)]{Fender04}
Fender, R.P.; Belloni, T.M.; Gallo, E. Towards a unified model for black hole X-ray binary jets. {\em Mon. Not. R. Astron. Soc.} {\bf 2004}, {\em 355}, 1105--1118.

\bibitem[Gallo et al. (2012)]{Gallo12}
Gallo, E.; Miller, B.; Fender, R. Assessing luminosity correlations via cluster analysis: Evidence for dual tracks in the radio/X-ray domain of black hole X-ray binaries. {\em Mon. Not. R. Astron. Soc.} {\bf 2012}, {\em 423}, 590--599.

\bibitem[Corbel et al.(2013)]{corbel13}
Corbel, S.; Coriat, M.; Brocksopp, C.; Tzioumis, A.K.; Fender, R.P.; Tomsick, J.A.; Buxton, M.M.; Bailyn, C.D. The `universal' radio/X-ray flux correlation: The case study of the black hole GX 339-4. {\em Mon. Not. R. Astron. Soc.} {\bf 2013}, {\em 428}, 2500--2515.

\bibitem[Tetarenko et al.(2021)]{Tetarenko21}
Tetarenko, A.J.; Casella, P.; Miller-Jones, J.C.A.; Sivakoff, G.R.; Paice, J.A.; Vincentelli, F.M.; Maccarone, T.J.; Gandhi, P.; Dhillon,~V.S.; Marsh, T.R.; et al. Measuring fundamental jet properties with multiwavelength fast timing of the black hole X-ray binary MAXI J1820+070. {\em Mon. Not. R. Astron. Soc.} {\bf 2021}, {\em 504}, 3862--3883.

\bibitem[Connors et al. (2019)]{Connors19}
Connors, R.M.T.; van Eijnatten, D.; Markoff, S.; Ceccobello, C.; Grinberg, V.; Heil, L.; Kantzas, D.; Lucchini, M.; Crumley, P. Combining timing characteristics with physical broad-band spectral modelling of black hole X-ray binary GX 339-4. {\em Mon. Not. R. Astron. Soc.} {\bf 2019}, {\em 485}, 3696--3714.

\bibitem[Motta et al.(2021)]{Motta21}
Motta, S.E.; Kajava, J.J.E.; Giustini, M.; Williams, D.R.A.; Del Santo, M.; Fender, R.; Green, D.A.; Heywood, I.; Rhodes, L.; Segreto,~A.; et al. Observations of a radio-bright, X-ray obscured GRS 1915+105. {\em Mon. Not. R. Astron. Soc.} {\bf 2021}, {\em 503}, 152--161.

\bibitem[Paragi et al. (2012)]{Paragi12}
Paragi, Z.; Vermeulen, R.; Spencer, R.E. SS433, Microquasars, and Other Transients. In Proceedings of the Meeting ``Resolving The Sky---Radio Interferometry: Past, Present and Future'', Manchester, UK, 17--20 April 2012; p. 28. Available online: \url{http://pos.sissa.it/cgi-bin/reader/conf.cgi?confid=163} (accessed on 6 September 2021).


\bibitem[Gallo et al.(2019)]{Gallo19}
Gallo, E.; Teague, R.; Plotkin, R.M.; Miller-Jones, J.C.A.; Russell, D.M.; Dincer, T.; Bailyn, C.; Maccarone, T.J.; Markoff, S.; Fender,~R.P. ALMA observations of A0620-00: Fresh clues on the nature of quiescent black hole X-ray binary jets. {\em Mon. Not. R. Astron. Soc.} {\bf 2019}, {\em 488}, 191--197.

\bibitem[Tremou et al.(2020)]{Tremou20}
Tremou, E.; Corbel, S.; Fender, R.P.; Woudt, P.A.; Miller-Jones, J.C.A.; Motta, S.E.; Heywood, I.; Armstrong, R.P.; Groot, P.; \mbox{Horesh, A.; et al.} Radio \& X-ray detections of GX 339-4 in quiescence using MeerKAT and Swift. {\em Mon. Not. R. Astron. Soc.} {\bf 2020}, {\em 493}, L132--L137.

\bibitem[Plotkin et al.(2017)]{Plotkin17}
Plotkin, R.M.; Bright, J.; Miller-Jones, J.C.A.; Shaw, A.W.; Tomsick, J.A.; Russell, T.D.; Zhang, G.-B.; Russell, D.M.; Fender, R.P.; Homan, J.; et al. Up and Down the Black Hole Radio/X-ray Correlation: The 2017 Mini-outbursts from Swift J1753.5-0127. {\em Astrophys. J.} {\bf 2017}, {\em 848}, 92.

\bibitem[Jonker et al. (2010)]{Jonker10}
\textls[-10]{Jonker, P.G.; Miller-Jones, J.; Homan, J.; Gallo, E.; Rupen, M.; Tomsick, J.; Fender, R.P.; Kaaret, P.; Steeghs, D.T.H.; Torres,~M.A.P.;~et~al. Following the 2008 outburst decay of the black hole candidate H 1743-322 in X-ray and radio. {\em Mon. Not. R. Astron. Soc.} {\bf 2010}, {\em 401}, 1255--1263.}

\bibitem[Coriat et al. (2011)]{Criat11}
Coriat, M.; Corbel, S.; Prat, L.; Miller-Jones, J.C.A.; Cseh, D.; Tzioumis, A.K.; Brocksopp, C.; Rodriguez, J.; Fender, R.P.; Sivakoff,~G.R. Radiatively efficient accreting black holes in the hard state: The case study of H1743-322. {\em Mon. Not. R. Astron. Soc.} {\bf 2011}, {\em 414}, 677--690.

\bibitem[Tomsick et al.(2015)]{Tomsick15}
Tomsick, J.A.; Rahoui, F.; Kolehmainen, M.; Miller-Jones, J.; F$\ddot{u}$rst, F.; Yamaoka, K.; Akitaya, H.; Corbel, S.; Coriat, M.; Done,~C.;~et~al. The Accreting Black Hole Swift J1753.5-0127 from Radio to Hard X-ray. {\em Astrophys. J.} {\bf 2015}, {\em 808}, 85.

\bibitem[Remillard \& McClintock(2006)]{Remillard06}
Remillard, R.A.; McClintock, J.E. X-ray Properties of Black-Hole Binaries. {\em Annu. Rev. Astron. Astrophys.} {\bf 2006}, {\em 44}, 49--92.

\bibitem[Plotkin et al. (2013)]{Plotkin13}
Plotkin, R.M.; Gallo, E.; Jonker, P.G. The X-ray Spectral Evolution of Galactic Black Hole X-ray Binaries toward Quiescence. {\em Astrophys. J.} {\bf 2013}, {\em 773}, id59.

\bibitem[Gallo et al.(2014)]{Gallo14}
Gallo, E.; Miller-Jones, J.C.A.; Russell, D.M.; Jonker, P.G.; Homan, J.; Plotkin, R.M.; Markoff, S.; Miller, B.P.; Corbel, S.; Fender,~R.P. The radio/X-ray domain of black hole X-ray binaries at the lowest radio luminosities. {\em Mon. Not. R. Astron. Soc.} {\bf 2014}, {\em 445}, 290--300.

\bibitem[Blandford \& K$\ddot{o}$nigl(1979)]{bland79}
Blandford, R.D.; K$\ddot{o}$nigl, A. Relativistic jets as compact radio sources. {\em Astrophys. J.} {\bf 1979}, {\em 232}, 34--48.

\bibitem[Fender (2001)]{Fender01}
Fender, R.P. Powerful jets from black hole X-ray binaries in low/hard X-ray states. {\em Mon. Not. R. Astron. Soc.} {\bf 2001}, {\em 322}, 31--42.

\bibitem[Heinz \& Sunyaev (2003)]{Heinz03}
Heinz, S.; Sunyaev, R.A. The non-linear dependence of flux on black hole mass and accretion rate in core-dominated jets. {\em Mon.~Not. R. Astron. Soc.} {\bf 2003}, {\em 343}, L59--L63.

\bibitem[Willott et al. (1999)]{willott99}
Willott, C.J.; Rawlings, S.; Blundell, K.M.; Lacy, M. The emission line-radio correlation for radio sources using the 7C Redshift Survey. {\em Mon. Not. R. Astron. Soc.} {\bf 1999}, {\em 309}, 1017-1033.

\bibitem[Godfrey \& Shabala (2013)]{Godfrey13}
Godfrey, L.E.H.; Shabala, S.S. AGN Jet Kinetic Power and the Energy Budget of Radio Galaxy Lobes. {\em Astrophys. J.} {\bf 2013}, {\em 767}, id12.

\bibitem[Yuan \& Narayan (2014)]{Yuan14}
Yuan, F.; Narayan, R. Hot Accretion Flows Around Black Holes. {\em Annu. Rev. Astron. Astrophys.} {\bf 2014}, {\em 52}, 529.

\bibitem[Shakura \& Sunyaev (1973)]{Shakura73}
Shakura, N.I.; Sunyaev, R.A. Black holes in binary systems. Observational appearance. {\em Astron. Astrophys.} {\bf 1973}, {\em 500}, 33--51.


\bibitem[Mirabel \&  Rodr\'iguez (1994)]{Mirabel1994}
Mirabel, I.F.; Rodr\'iguez, L.F. A superluminal source in the Galaxy. {\em Nature} {\bf 1994}, {\em 371}, 46--48.

\bibitem[Novikov \& Thorne (1973)]{NovikovT}
Novikov, I.D.; Thorne, K.S. Astrophysics in black holes. In {\em Black Holes (Les Astres OccLus)}; DeWitt, C., DeWitt, B., Eds.; Gordon and Breach: New York, NY, USA, 1973; pp. 343--350. 

\bibitem[Narayan et al. (2021)]{Narayan21}
Narayan, R.; Chael, A.; Chatterjee, K.; Ricarte, A.; Curd, B. Jets in Magnetically Arrested Hot Accretion Flows: Geometry, Power and Black Hole Spindown. {\em arXiv} {\bf 2021}, arXiv:2108.12380.

\bibitem[Espinasse \& Fender (2018)]{Espinasse18}
Espinasse, M.; Fender, R. Spectral differences between the jets in `radio-loud' and `radio-quiet' hard-state black hole binaries. {\em Mon. Not. R. Astron. Soc.} {\bf 2018}, {\em 473}, 4122--4129.

\bibitem[Yuan et al. (2005)]{Yuan05}
Yuan, F.; Cui, W.; Narayan, R. An Accretion-Jet Model for Black Hole Binaries: Interpreting the Spectral and Timing Features of XTE J1118+480. {\em Astrophys. J.} {\bf 2005}, {\em 620}, 905--914.

\bibitem[Narayan, Mahadevan \& Quataert (1998)]{narayan1998}
Narayan, R.; Mahadevan, R.; Quataert, E. Advection-Dominated Accretion around Black Holes. In {\em Theory of Black Hole Accretion Disks}; Abramowicz, M.A., Bjornsson, G., Pringle, J.E., Eds.; Cambridge University Press: Cambridge, UK, 1998; pp. 148--182.

\bibitem[Narayan \& Yi (1994)]{Narayan94}
Narayan, R.; Yi, I. Advection-dominated Accretion: A Self-similar Solution. {\em Astrophys. J.} {\bf 1994}, {\em 428}, L13.

\bibitem[Xie \& Yuan (2012)]{Xie12}
Xie, F.-G.; Yuan, F. Radiative efficiency of hot accretion flows. {\em Mon. Not. R. Astron. Soc.} {\bf  2012}, {\em 427}, 1580--1586.

\bibitem[Xie \& Yuan (2016)]{XieY16}
Xie, F.-G. ; Yuan, F. Interpreting the radio/X-ray correlation of black hole X-ray binaries based on the accretion-jet model. {\em Mon.~Not. R. Astron. Soc.} {\bf 2016}, {\em 456}, 4377--4383.

\bibitem[Meyer-Hofmeister \& Meyer (2014)]{Meyer-Hofmeister2014}
Meyer-Hofmeister, E.; Meyer, F. The relation between radio and X-ray luminosity of black hole binaries: Affected by inner cool disks? {\em Astron. Astrophys.} {\bf 2014}, {\em 562}, A142.

\bibitem[Laor \& Behar (2008)]{Laor08}
Laor, A.; Behar, E. On the origin of radio emission in radio-quiet quasars. {\em Mon. Not. R. Astron. Soc.} {\bf 2008}, {\em 390}, 847--862.

\bibitem[Wallace \& Pe'er (2021)]{Wallace21}
Wallace, J.; Pe'er, A. An Observational Signature of Sub-equipartition Magnetic Fields in the Spectra of Black Hole Binaries. {\em Astrophys. J.} {\bf 2021}, {\em 916}, 63.

\bibitem[Dong et al. (2021)]{dong21}
Dong, A.-J., Liu, C.; Ge, K.; Liu, X.; Zhi, Q.-J.; You, Z.-Y. A Study on the Hysteresis Effect and Spectral Evolution in the Mini-Outbursts of Black Hole X-ray Binary XTE J1550-564. {\em Front. Astron. Space Sci.} {\bf 2021}, {\em 8}, 37.

\bibitem[Xie et al. (2020)]{Xie20}
Xie, F.-G.; Yan, Z.; Wu, Z. Radio/X-ray Correlation in the Mini-outbursts of Black Hole X-ray Transient GRS 1739-278. {\em Astrophys.~J.} {\bf 2020}, {\em 891}, id31.

\bibitem[Carotenuto et al. (2021)]{Carotenuto21}
Carotenuto, F.; Corbel, S.; Tremou, E.; Russell, T.D.; Tzioumis, A.; Fender, R.P.; Woudt, P.A.; Motta, S.E.; Miller-Jones, J.C.A.; Tetarenko, A.J.; et al. The hybrid radio/X-ray correlation of the black hole transient MAXI J1348-630. {\em Mon. Not. R. Astron. Soc.} {\bf 2021}, {\em 505}, L58--L63.

\bibitem[Steiner et al. (2013)]{Steiner13}
Steiner, James F.; McClintock, J.E.; Narayan, R. Jet Power and Black Hole Spin: Testing an Empirical Relationship and Using it to Predict the Spins of Six Black Holes. {\em Astrophys. J.} {\bf 2013}, {\em 762}, 104.

\bibitem[Blandford \& Znajek (1977)]{bland77}
Blandford, R.D.; Znajek, R.L. Electromagnetic extraction of energy from Kerr black holes. {\em Mon. Not. R. Astron. Soc.} {\bf 1977} {\em 179}, 433.

\bibitem[Blandford \& Payne (1982)]{bland82}
Blandford, R.D.; Payne, D.G. Hydromagnetic flows from accretion disks and the production of radio jets. {\em Mon. Not. R. Astron. Soc.} {\bf 1982}, {\em 199}, 883.

\bibitem[Reynolds et al. (2021)]{Reynolds21}
Reynolds, C.S. Observational Constraints on Black Hole Spin. {\em Annu. Rev. Astron. Astrophys.} {\bf 2021}, {\em 59}, 117--154.

\bibitem[Gou et al. (2014)]{Gou14}
Gou, L.; McClintock, J.E.; Remillard, R.A.; Steiner, J.F.; Reid, M.J.; Orosz, J.A.; Narayan, R.; Hanke, M.; Garcia, J. Confirmation via the Continuum-fitting Method that the Spin of the Black Hole in Cygnus X-1 Is Extreme. {\em Astrophys. J.} {\bf 2014}, {\em 790}, 29.

\bibitem[Merloni, Heinz \& di Matteo (2003)]{Merloni03}
Merloni, A.; Heinz, S.; di Matteo, T. A Fundamental Plane of black hole activity. {\em Mon. Not. R. Astron. Soc.} {\bf 2003}, {\em 345}, 1057--1076.

\bibitem[Falcke, K\"ording \& Markoff (2004)]{Falcke04}
Falcke, H.; K\''ording, E.; Markoff, S. A scheme to unify low-power accreting black holes. Jet-dominated accretion flows and the radio/X-ray correlation. {\em Astron. Astrophys.} {\bf 2004}, {\em 414}, 895--903.

\bibitem[Liu et al. (2016)]{liu16}
Liu, X., Han, Z.H., Zhang, Z. The physical fundamental plane of black hole activity: Revisited. {\em Astrophys. Space Sci.} {\bf 2016}, {\em 361}, 9.

\bibitem[Liu et al. (2020)]{liu20}
Liu, X.; Chang, N.; Han, Z.; Wang, X. The Jet-Disk Coupling of Seyfer Galaxies from a Complete Hard X-ray Sample. {\em Universe} {\bf 2020}, {\em 6}, 68.

\bibitem[Baldi et al. (2021)]{Baldi2021}
\textls[-10]{Baldi, R.D.; Williams, D.R.A.; Beswick, R.J.; McHardy, I.; Dullo, B.T.; Knapen, J.H.; Zanisi, L.; Argo, M.K.; Aalto, S.; Alberdi, A.; et~al}. LeMMINGs. III. The e-MERLIN Legacy Survey of the Palomar sample. Exploring the origin of nuclear radio emission in active and inactive galaxies through the [O III]---Radio connection. {\em Mon. Not. R. Astron. Soc.} {\bf 2021}, {\em 508}, 2019--2038.

\bibitem[Feng et al. (2021)]{Feng21}
Feng, J.J.; Cao, X.W.; Li, J.W.; Gu, W.M. A Magnetic Disk-outflow Model for Changing Look Active Galactic Nuclei. {\em Astrophys. J.} {\bf 2021}, {\em 916}, 61.

\bibitem[Yuan \& Zdziarsk (2004)]{YuanZ04}
Yuan, F.; Zdziarski, A.A. Luminous hot accretion flows: The origin of X-ray emission from Seyfert galaxies and black hole binaries. {\em Mon. Not. R. Astron. Soc.} {\bf 2004}, {\em 354}, 953--960.

\bibitem[Motta et al. (2018)]{motta2018}
Motta, S.E.; Casella, P.; Fender, R.P. Radio-loudness in black hole transients: Evidence for an inclination effect. {\em Mon. Not. R. Astron. Soc.} {\bf 2018}, {\em 478}, 5159--5173.

\bibitem[Narayan et al. (2003)]{Narayan03}
Narayan, R.; Igumenshchev, I.V.A.; Abramowicz, M.A.J. Magnetically Arrested Disk: An Energetically Efficient Accretion Flow. {\em Pub. Astron. Soc. Japan} {\bf 2003}, {\em 55}, L69--L72.

\bibitem[Tchekhovskoy et al. (2011)]{Tchekhovskoy11}
Tchekhovskoy, A.; Narayan, R.; McKinney, J.C. Efficient generation of jets from magnetically arrested accretion on a rapidly spinning black hole. {\em Mon. Not. R. Astron. Soc.} {\bf 2011}, {\em 418}, L79--L83.

\bibitem[Tchekhovskoy \& McKinney (2012)]{Tchekhovskoy12}
Tchekhovskoy, A.; McKinney, J.C. Prograde and retrograde black holes: Whose jet is more powerful? {\em Mon. Not. R. Astron. Soc.} {\bf 2012}, {\em 423}, L55--L59.

\bibitem[Debnath et al. (2021)]{Debnath21}
Debnath, D.; Chatterjee, K.; Chatterjee, D.; Jana, A.; Chakrabarti, S.K. Jet properties of XTE J1752-223 during its 2009--2010 outburst. {\em Mon. Not. R. Astron. Soc.} {\bf 2021}, {\em 504}, 4242--4251.






\end{thebibliography}
\end{document}